\documentclass[aps,preprint]{revtex4}%
\usepackage{amsfonts}
\usepackage{amsmath}
\usepackage{amssymb}
\usepackage{graphicx}%
\begin{document}
\preprint{HEP/123-qed}
\title[Short title for running header]{Atomic entanglement generation with reduced decoherence via four-wave mixing}
\author{C. Genes}
\affiliation{Michigan Center for Theoretical Physics, FOCUS Center, and Physics Department,
University of Michigan, Ann Arbor 48109-1040, USA}
\author{P. R. Berman}
\affiliation{Michigan Center for Theoretical Physics, FOCUS Center, and Physics Department,
University of Michigan, Ann Arbor 48109-1040, USA}

\begin{abstract}
In most proposals for the generation of entanglement in large ensembles of
atoms via projective measurements, the interaction with the vacuum is
responsible for both the generation of the signal that is detected and the
spin depolarization or decoherence. In consequence, one has to usually work in
a regime where the information aquisition via detection is sufficiently slow
(weak measurement regime) such as not to strongly disturb the system. We
propose here a four-wave mixing scheme where, owing to the pumping of the
atomic system into a dark state, the polarization of the ensemble is not
critically affected by spontaneous emission, thus allowing one to work in a
strong measurement regime.

\end{abstract}
\volumeyear{year}
\volumenumber{number}
\issuenumber{number}
\eid{identifier}
\date[Date text]{date}
\received[Received text]{date}

\revised[Revised text]{date}

\accepted[Accepted text]{date}

\published[Published text]{date}

\startpage{1}
\maketitle

\section{Introduction}

Considerable attention has been given in recent years to the generation of
multi-particle entanglement in large ensembles of long-lived atomic spins.
Some authors have proposed to achieve this goal using interactions between
atoms and light, where a quantum state exchange can take place \cite{polzik
97, molmer-polzik 2000, genes 2003, dantan 2004, dantan 2005}, leading to the
preparation of a desired collective spin state. A different set of proposals
make use of an appropriate measurement on one of the field's observables that
leads to the collapse of the ensemble onto the desired entangled spin state.
In general, these schemes fall into two categories: conditional \cite{kuzmich
98, takahashi, kuzmich 2000, molmer 2002, duan-kimble 2003, sorensen-molmer
2003, massar-polzik} and deterministic \cite{wiseman1, wiseman2, mabuchi
2004}. In a conditional entanglement generation scheme, the prepared state of
the atoms is conditioned by the outcome of the measurement on the field state.
The random character of the state resulting from the measurement back-action
can be removed if one performs a continuous quantum nondemolition measurement
of a spin observable and adding a feedback loop for \textit{a posteriori}
quantum state correction based on the detection outcome. In this way a
deterministic (unconditional) quantum state preparation scheme can be
realized, where the uncertainty in the final state is removed.

In general, multilevel atoms with long-lived ground substates are used, where
two of the ground sublevels form an effective two-level atom. The manipulation
of the collective atomic spin, obtained by summation over the individual spins
associated with each atom in the medium, is achieved by driving ground-excited
state transitions using classical or quantized radiation fields. As a result
of the interactions, a signal field is generated that reflects some quantum
mechanical fluctuations in the atomic ensemble; its detection can give
information about the atomic ensemble state. In the off-resonance regime, an
effective Hamiltonian can be found with a coupling between atoms and signal
field that is proportional to the strength of the atom's coupling to the
vacuum and also to the amplitude of the driving field. An increase in the
control field's amplitude, therefore, would seem to allow one to generate
optimal entanglement. However, the downside of using coupling through excited
levels is that spontaneous emission comes into play, leading to a rapid
decoherence of the system. In consequence, it is necessary to limit
spontaneous emission losses to a small value, which forces one to work in a
regime of weak coupling of the atomic system to the field system to be
measured, and only weak entanglement can be obtained.

The competition between spontaneous emission and measurement strength is best
illustrated in the case of spin squeezed state \cite{spin squeezing concept
papers} generation. The challenge there is to reduce fluctuations in a spin
component orthogonal to the mean spin, while keeping the average spin length
large. The measurement strength controls the reduction of fluctuations, while
spontaneous emission leads to a diminishing of the spin length. In a recent
publication \cite{berman-genes}, it has been shown that, for the case of a
pencil-shaped medium with Fresnel number close to unity, in the regime of
small decay, optimal results are limited by the resonant optical depth of the sample.

Despite the interplay between measurement strength and spontaneous decay, it
is possible to imagine a situation in which spontaneous decay does not limit
the value of the measurement coupling strength. In one such scenario, the
system is prepared in a dark state which is preserved during the interaction
by a convenient choice of driving fields. This is the situation presented in
this paper, where a quantized signal field, generated via four-wave mixing in
a double $\Lambda$ atomic system in a pencil-shaped medium, is entangled with
the collective atomic state. The generated signal pulse reflects fluctuations
in the population difference between the two ground substates. The measurement
on the signal field photon number can, therefore, give information on the $z$
component of the atomic spin, projecting the system into either a spin
squeezed or Schrodinger cat state. At the same time, as opposed to the
situation illustrated in \cite{berman-genes}, the collective $x$ polarized
spin state in which the system is initially prepared, is very nearly a dark
state for this combination of fields and is subject to minimal decay. No
severe limitations on the measurement coupling strength are therefore
necessary, and results similar to the ideal case presented in
\cite{berman-genes} are obtained. With the assumption of perfect detection,
the Heisenberg limit is the ultimate limitation to the squeezing parameter.

The paper is organized as follows: in Sec. II the proposed scheme is described
and an analytical expression for the atomic operator giving rise to the signal
field is obtained; spontaneous emission effects are also discussed. In Sec.
III, an expression for the signal field amplitude operator is obtained, which
is shown to reflect atomic population fluctuations. An effective Hamiltonian
necessary for a wave function description of the problem is derived in Sec.
IV, while in Sec. V the generation of entanglement via conditional measurement
of the signal field is discussed. Some conclusions are presented in Sec. VI.

\section{Scheme and method}

\begin{figure}
[ptb]
\begin{center}
\includegraphics[
height=3.0632in,
width=5.0816in
]%
{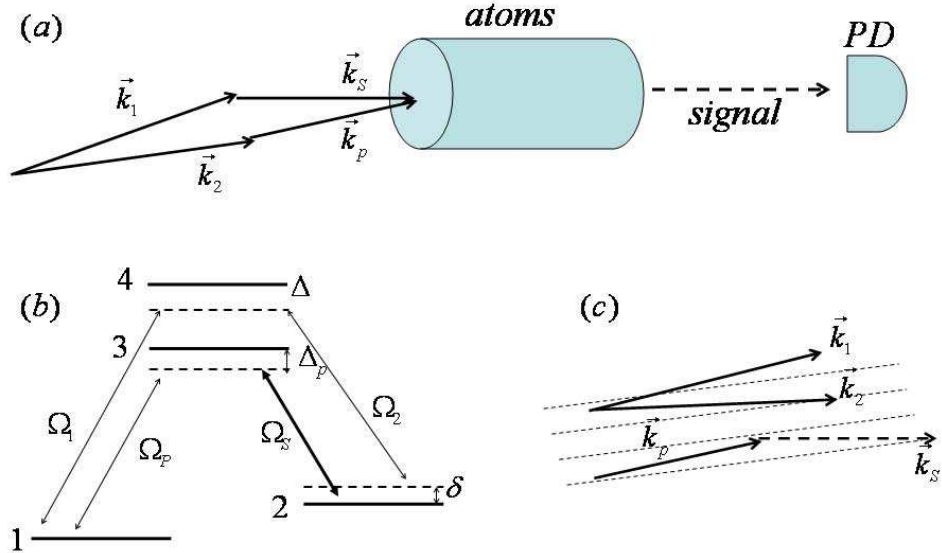}%
\caption{(a) The three classical pulses indexed by $1$, $2~$and $p~$are
incident on the sample at an angle giving rise to a signal field that
propagates along the cylinder's axis. \ Detection of photon number takes place
at the photodetector (PD). (b) Illustration of the internal structure of an
atom as a double $\Lambda~$scheme. (c) The signal field is respresented as a
reflection of the probe ($p$) off a spatial grating generated by the pumps
($1$~and $2$) }%
\label{scheme}%
\end{center}
\end{figure}

A pencil-shaped atomic medium aligned along the $z$ axis (left end situated at
$z=0$), with transverse area $A$, length $L$, and density $n_{a}$ is
considered. The internal structure of an atom [Fig. \ref{scheme}(b)] is a
double $\Lambda$ scheme with ground levels $1$ and $2$ and excited levels $3$
and $4$. Three classical laser pulses having duration $T\gg L/c$ (where $c$ is
the speed of light), wave vectors $\mathbf{k}_{1}$, $\mathbf{k}_{2}$ (pumps)
and $\mathbf{k}_{p}$ (probe), and frequencies $\Omega_{1}$, $\Omega_{2}$ and
$\Omega_{p}$ are simultaneously shined on the atoms. We consider an
off-resonant regime with one-photon detuning $\Delta$ for the pump fields,
one-photon detuning $\Delta_{p}$ for the probe field, and two-photon Raman
detuning $\delta$ [see Fig. \ref{scheme}(b)]. The three classical waves mix
inside the atomic medium to generate a multitude of secondary waves; the one
which propagates along the positive $z$ direction [see Fig. \ref{scheme}(a)]
and which can be viewed as a reflection of the first pump wave off a spatial
grating produced by the second pump and the probe is of interest and denoted
as the signal wave. This wave is radiated on the $2\rightarrow3$ transition,
and has frequency $\Omega_{s}=\Omega_{p}-(\Omega_{1}-\Omega_{2})$ and phase
matched wave vector $\mathbf{k}_{s}=\left(  \Omega_{s}/c\right)  \widehat
{z}=\mathbf{k}_{p}-(\mathbf{k}_{1}-\mathbf{k}_{2})$. A few methods can be
employed to separate the signal wave from the probe and pumps. The simplest
one [illustrated in Fig. \ref{scheme}(a)] requires an increase of the angles
made by the three primary wave with the $z$ axis sufficient to provide a clear
angular resolution. The other two involve both the use of a polarization beam
splitter (to distinguish between probe, pump $1$ and signal) or a spectral
filter (that can distinguish between pump $2$ and signal). Finally, a
photodetector (PD) is used to detect the photon number of the signal field.%
In the following, we consider the dynamics of an atom located inside the
sample, at position $\mathbf{r}$.

The three incoming pulses are represented as $c~$number traveling waves
\begin{equation}
\mathbf{E}^{(i)}(\mathbf{r},t)=\frac{1}{2}\left[  E_{i}(\mathbf{r}%
,t)e^{i\left(  \mathbf{k}_{i}\mathbf{\cdot r}-\Omega_{i}t\right)  }%
+E_{i}^{\ast}(\mathbf{r},t)e^{-i\left(  \mathbf{k}_{i}\mathbf{\cdot r}%
-\Omega_{i}t\right)  }\right]  \widehat{\epsilon}_{i}\text{,}%
\end{equation}
for $i=1,2,p$, with polarization unit vectors $\widehat{\epsilon}_{i}$. The
matrix elements of the atomic dipole $\mathbf{d}$ between atomic states $j$
and $j^{\prime}$ (along $\widehat{\epsilon}_{i}$) is denoted by $d_{jj^{\prime
}}=\left\langle j\left\vert \mathbf{d\cdot}\widehat{\epsilon}_{i}\right\vert
j^{\prime}\right\rangle $. The interaction between fields and atom is
described by the (spatially dependent) Rabi frequencies
\begin{subequations}
\begin{align}
\Lambda_{1}(\mathbf{r},t)  &  =\frac{d_{14}^{\ast}E_{1}(\mathbf{r},t)}{2\hbar
}e^{i\mathbf{k}_{i}\mathbf{\cdot r}}=\chi_{1}(\mathbf{r},t)e^{i\mathbf{k}%
_{i}\mathbf{\cdot r}},\\
\Lambda_{2}(\mathbf{r},t)  &  =\frac{d_{24}^{\ast}E_{2}(\mathbf{r},t)}{2\hbar
}e^{i\mathbf{k}_{2}\mathbf{\cdot r}}=\chi_{2}(\mathbf{r},t)e^{i\mathbf{k}%
_{2}\mathbf{\cdot r}},\\
\Lambda_{p}(\mathbf{r},t)  &  =\frac{d_{13}^{\ast}E_{p}(\mathbf{r},t)}{2\hbar
}e^{i\mathbf{k}_{p}\mathbf{\cdot r}}=\chi_{p}(\mathbf{r},t)e^{i\mathbf{k}%
_{p}\mathbf{\cdot r}}.
\end{align}
Assuming that the pump and probe pulses are long [$cT\gg L$] and wide
(transverse area larger than $A$), we can neglect the slow variation with
$\mathbf{r}$ in $\chi_{j}(\mathbf{r},t)$; the field envelope is replaced with
its value at $\mathbf{r}=0$, and $\chi_{j}(0,t)$ is replaced with $\chi
_{j}(t)$.

Atomic operators are defined as $\sigma_{ii}=\left\vert i\right\rangle
\left\langle i\right\vert ~$(population operators), for $i=1,2,3,4$, and
$\sigma_{ij}=\left\vert i\right\rangle \left\langle j\right\vert
~$(coherences), for $i,j=1,2,3,4$ but with $i\neq j$. Ignoring the coupling to
the vacuum for the moment, the Hamiltonian that describes the evolution of the
atom driven by the three fields is a sum of the free Hamiltonian ($H_{0}$) and
the classical field-atom interaction Hamiltonian ($V_{c}$)%
\end{subequations}
\begin{subequations}
\begin{align}
H_{0}  &  =\hbar\omega_{21}\sigma_{22}+\hbar\omega_{31}\sigma_{33}+\hbar
\omega_{41}\sigma_{44},\\
V_{c}  &  =-\hbar\left[  \Lambda_{1}(\mathbf{r},t)e^{-i\Omega_{1}t}\sigma
_{41}+\Lambda_{1}^{\ast}(\mathbf{r},t)e^{i\Omega_{1}t}\sigma_{14}\right]  -\\
&  -\hbar\left[  \Lambda_{2}(\mathbf{r},t)e^{-i\Omega_{2}t}\sigma_{42}%
+\Lambda_{2}^{\ast}(\mathbf{r},t)e^{i\Omega_{2}t}\sigma_{24}\right]
-\nonumber\\
&  -\hbar\left[  \Lambda_{p}(\mathbf{r},t)e^{-i\Omega_{p}t}\sigma_{31}%
+\Lambda_{p}^{\ast}(\mathbf{r},t)e^{i\Omega_{p}t}\sigma_{13}\right]
.\nonumber
\end{align}
In the Heisenberg picture, the rapid time variation in the coherences is
removed: $\sigma_{14}=$ $\widetilde{\sigma}_{14}e^{-i\Omega_{1}t}$,
$\sigma_{24}=\widetilde{\sigma}_{24}e^{-i\Omega_{2}t}$, $\sigma_{13}%
=\widetilde{\sigma}_{13}e^{-i\Omega_{p}t}$, $\sigma_{12}=\widetilde{\sigma
}_{12}e^{-i(\Omega_{1}-\Omega_{2})t}$, $\widetilde{\sigma}_{34}=\widetilde
{\sigma}_{34}e^{-i(\Omega_{1}-\Omega_{p})t}$ and $\sigma_{23}=\widetilde
{\sigma}_{23}e^{-i[\Omega_{p}-(\Omega_{1}-\Omega_{2})]t}$. Equations of motion
for the slowly varying atomic operators ($\widetilde{\sigma}_{ij}$) are
obtained (dropping the tildes) as
\end{subequations}
\begin{subequations}
\label{Eqs. motion}%
\begin{align}
\frac{d}{dt}\sigma_{11}(t)  &  =-i\Lambda_{1}(\mathbf{r},t)\sigma
_{41}+i\Lambda_{1}^{\ast}(\mathbf{r},t)\sigma_{14}-i\Lambda_{p}(\mathbf{r}%
,t)\sigma_{31}+i\Lambda_{p}^{\ast}(\mathbf{r},t)\sigma_{13},\\
\frac{d}{dt}\sigma_{22}(t)  &  =-i\Lambda_{2}(\mathbf{r},t)\sigma
_{42}+i\Lambda_{2}^{\ast}(\mathbf{r},t)\sigma_{24},\\
\frac{d}{dt}\sigma_{33}(t)  &  =i\Lambda_{p}(\mathbf{r},t)\sigma_{31}%
-i\Lambda_{p}^{\ast}(\mathbf{r},t)\sigma_{13},\\
\frac{d}{dt}\sigma_{44}(t)  &  =i\Lambda_{1}(\mathbf{r},t)\sigma_{41}%
-i\Lambda_{1}^{\ast}(\mathbf{r},t)\sigma_{14}+i\Lambda_{2}(\mathbf{r}%
,t)\sigma_{42}-i\Lambda_{2}^{\ast}(\mathbf{r},t)\sigma_{24},\\
\frac{d}{dt}\sigma_{14}(t)  &  =-i\Delta\sigma_{14}+i\Lambda_{1}%
(\mathbf{r},t)\left[  \sigma_{11}-\sigma_{44}\right]  +i\Lambda_{2}%
(\mathbf{r},t)\sigma_{12}-i\Lambda_{p}(\mathbf{r},t)\sigma_{34},\\
\frac{d}{dt}\sigma_{24}(t)  &  =-i\left(  \Delta+\delta\right)  \sigma
_{24}+i\Lambda_{2}(\mathbf{r},t)\left[  \sigma_{22}-\sigma_{44}\right]
+i\Lambda_{1}(\mathbf{r},t)\sigma_{21},\\
\frac{d}{dt}\sigma_{13}(t)  &  =-i\Delta_{p}\sigma_{13}+i\Lambda
_{p}(\mathbf{r},t)\left[  \sigma_{11}-\sigma_{33}\right]  -i\Lambda
_{1}(\mathbf{r},t)\sigma_{43},\\
\frac{d}{dt}\sigma_{34}(t)  &  =-i\left(  \Delta-\Delta_{p}\right)
\sigma_{34}+i\Lambda_{1}(\mathbf{r},t)\sigma_{31}+i\Lambda_{2}(\mathbf{r}%
,t)\sigma_{32}-i\Lambda_{p}^{\ast}(\mathbf{r},t)\sigma_{14},\\
\frac{d}{dt}\sigma_{23}(t)  &  =-i\left(  \Delta_{p}+\delta\right)
\sigma_{23}+i\Lambda_{p}(\mathbf{r},t)\sigma_{21}-i\Lambda_{2}(\mathbf{r}%
,t)\sigma_{43},\\
\frac{d}{dt}\sigma_{12}(t)  &  =i\delta\sigma_{12}-i\Lambda_{1}(\mathbf{r}%
,t)\sigma_{42}+i\Lambda_{2}^{\ast}(\mathbf{r},t)\sigma_{14}-i\Lambda
_{p}(\mathbf{r},t)\sigma_{32}.
\end{align}
We are interested in the time and spatial modulation of the $\sigma_{23}$
coherence, which is responsible with the generation of the signal. The rapid
time variation of $\sigma_{23}$, at frequency $\Omega_{s}=\Omega_{p}%
-(\Omega_{1}-\Omega_{2})$ has already been eliminated. The terms giving a
spatial modulation at the correct, phase-matched wave vector $\mathbf{k}%
_{s}=\mathbf{k}_{p}-(\mathbf{k}_{1}-\mathbf{k}_{2})$, have to be identified in
Eq. (\ref{Eqs. motion}(i)).

We proceed to describe the dynamics of the atom by using a perturbative
approach to solve Eqs. (\ref{Eqs. motion}), in which terms up to the third
order in $\chi/\Delta_{i}$ ($\chi=\chi_{1},\chi_{2},\chi_{3}$ and $\Delta
_{i}=\Delta,\Delta_{p},\delta$) are kept. We start with an $x$ polarized state
of the two-level system formed by the ground sublevels $1$ and $2$, where
$\sigma_{11}(0)=\sigma_{11}^{0}=1/2$, $\sigma_{22}(0)=\sigma_{22}^{0}=1/2$,
$\sigma_{12}(0)=\sigma_{12}^{0}=\sigma_{21}(0)=\sigma_{21}^{0}=-1/2$. Owing to
the assumptions of large detunings and long duration pulses, the upper state
population operators $\sigma_{ii}$ can be neglected. The coherences between
ground and excited states adiabatically follow the fields. Assuming $\delta
\ll\Delta,\Delta_{p}$, the first order solution of Eqs. (\ref{Eqs. motion}%
e,f,g,i) are given by:
\end{subequations}
\begin{subequations}
\begin{align}
\sigma_{14}^{(1)}(t)  &  \simeq\frac{\Lambda_{1}(\mathbf{r},t)}{\Delta}%
\sigma_{11}^{0}+\frac{\Lambda_{2}(\mathbf{r},t)}{\Delta}\sigma_{12}^{0},\\
\sigma_{24}^{(1)}(t)  &  \simeq\frac{\Lambda_{2}(\mathbf{r},t)}{\Delta}%
\sigma_{22}^{0}+\frac{\Lambda_{1}(\mathbf{r},t)}{\Delta}\sigma_{21}^{0},\\
\sigma_{13}^{(1)}(t)  &  \simeq\frac{\Lambda_{p}(\mathbf{r},t)}{\Delta_{p}%
}\sigma_{11}^{0},\\
\sigma_{23}^{(1)}(t)  &  \simeq\frac{\Lambda_{p}(\mathbf{r},t)}{\Delta_{p}%
}\sigma_{21}^{0}.
\end{align}
To second order, as observed before, the excited state populations have
derivatives which are identically zero; however, the ground states rate
equations are
\end{subequations}
\begin{subequations}
\begin{align}
\frac{d}{dt}\sigma_{11}^{(2)}(t)  &  \simeq i\left[  \frac{\Lambda_{1}^{\ast
}(\mathbf{r},t)\Lambda_{2}(\mathbf{r},t)}{\Delta}\sigma_{12}^{0}-\frac
{\Lambda_{1}(\mathbf{r},t)\Lambda_{2}^{\ast}(\mathbf{r},t)}{\Delta}\sigma
_{21}^{0}\right]  ,\\
\frac{d}{dt}\sigma_{22}^{(2)}(t)  &  =-\frac{d}{dt}\sigma_{11}^{(2)}(t).
\end{align}

An important observation can be made at this point. The system formed by the
two ground sublevels is driven by an effective field with a Rabi frequency
$[\chi_{1}^{\ast}(t)\chi_{2}(t)/\Delta]$ multiplied by a spatial phase,
dependent on the atom's position inside the medium: $e^{-i(\mathbf{k}%
_{1}-\mathbf{k}_{2})\mathbf{\cdot r}}$. In the copropagating pumps geometry,
this spatial modulation is negligibly small over the length of the medium
since $\left\vert \mathbf{k}_{1}-\mathbf{k}_{2}\right\vert L\simeq\omega
_{21}L/c\ll1$. In addition, assuming $\chi_{1}(t)$ and $\chi_{2}(t)$ are real,
the resulting effective Rabi frequency is real. The $\overset{\cdot}{\sigma
}_{11}^{(2)}$ is, in consequence, vanishing for any atom inside the medium,
independent on its location. Also, owing to the fact that equal the ground
state populations are equal, the $\overset{\cdot}{\sigma}_{12}$vanishes to
this order as well; in consequence the system stays in a dark state with equal
populations and coherence along the $x$ axis.

The two coherences, $\sigma_{34}$ and $\sigma_{12}$, that act as sources in
Eq. (\ref{Eqs. motion}i), are not driven directly by the field; their change
from initial values is a second order contribution. They can be derived from
Eqs. (\ref{Eqs. motion}h,j), which give
\end{subequations}
\begin{subequations}
\label{second order}%
\begin{align}
\sigma_{34}^{(2)}(t)  &  \simeq\frac{\Lambda_{p}^{\ast}(\mathbf{r},t)}%
{\Delta\Delta_{p}}\left[  \Lambda_{1}(\mathbf{r},t)\sigma_{11}^{0}+\Lambda
_{2}(\mathbf{r},t)\sigma_{12}^{0}\right]  ,\\
\sigma_{12}^{(2)}(t)  &  =\sigma_{12}^{0}+\frac{\Lambda_{1}(\mathbf{r}%
,t)\Lambda_{2}^{\ast}(\mathbf{r},t)}{\Delta\delta}\left[  \sigma_{11}%
^{0}-\sigma_{22}^{0}\right]  .
\end{align}
We are now in position to evaluate the third order approximation of
$\sigma_{23}$. Replacing Eqs. (\ref{second order}) in Eq. (\ref{Eqs. motion}%
i), one obtains
\end{subequations}
\begin{align}
\sigma_{23}^{(3)}(t)  &  \simeq\frac{\Lambda_{p}(\mathbf{r},t)}{\Delta_{p}%
}\sigma_{21}^{0}+\frac{\Lambda_{1}^{\ast}(\mathbf{r},t)\Lambda_{2}%
(\mathbf{r},t)\Lambda_{p}(\mathbf{r},t)}{\Delta\Delta_{p}\delta}\left[
\sigma_{11}^{0}-\sigma_{22}^{0}\right]  -\label{sigma23}\\
&  -\frac{\Lambda_{1}^{\ast}(\mathbf{r},t)\Lambda_{2}(\mathbf{r},t)\Lambda
_{p}(\mathbf{r},t)}{\Delta\Delta_{p}^{2}}\sigma_{11}^{0}-\frac{\left\vert
\Lambda_{2}(\mathbf{r},t)\right\vert ^{2}\Lambda_{p}(\mathbf{r},t)}%
{\Delta\Delta_{p}^{2}}\sigma_{21}^{0}.\nonumber
\end{align}
Two of the terms in the above expression [first and fourth in the right hand
side of Eq. (\ref{sigma23})] describe a field propagating in the direction of
$\mathbf{k}_{p}$, which is not phase matched [$k_{p}\neq\Omega_{s}/c$]. Both
other terms give rise to a phase matched signal field; however, in the limit
$\delta\ll\Delta_{p}$ the third term is negligible compared to the second one,
and is dropped. With the notation $\sigma_{z}=\left(  \sigma_{22}-\sigma
_{11}\right)  /2$, the expression of $\mathbf{\sigma}_{23}$ can be simplified
\begin{equation}
\sigma_{23}(t)=\frac{\Lambda_{1}^{\ast}(\mathbf{r},t)\Lambda_{2}%
(\mathbf{r},t)\Lambda_{p}(\mathbf{r},t)}{\Delta\Delta_{p}\delta}\left[
\sigma_{11}-\sigma_{22}\right]  . \label{sigma23final}%
\end{equation}

The analysis is not complete before the role of spontaneous decay is properly
identified. The conditions imposed on the fields guarantee that the effective
driving pulse doesn't remove the system from the dark state; spontaneous decay
can still destroy the coherence, as is the case in \cite{berman-genes}.
However, we start by making an observation on a simple system of a $\Lambda$
atom driven by two equal fields, in which, on each of the two transitions,
Raman and Rayleigh scatterings cancel each other, leading to a state not
affected by spontaneous emission. Our case is similar to this, although not
completely identical. In the absence of the probe field, an initially balanced
state (equal populations in the ground substates) with $-1/2$ coherence, would
be preserved by choosing equal amplitude pumps. The probe field provides an
imbalance in the system, which can be compensated by choosing a field strength
on the $2\rightarrow4$ transition larger than the one on the $1\rightarrow4$
transition by an amount that cancels the effect of the probe. This condition
is $\gamma^{\prime}(\chi_{1}^{2}-\chi_{2}^{2})/\Delta^{2}=\gamma\chi_{p}%
^{2}(t)/\Delta_{p}^{2}$ and is obtained by imposing the condition that the
decay terms in the rate equations for $\sigma_{11}(t)$ and $\sigma_{22}(t)$
are identically zero. In addition, the coherence between the ground sublevels
follows the fields with a slowly varying value of $-(\chi_{1}/\chi_{2})/2$. By
limiting the intensity of the probe field to small values compared to the pump
field intensities, this coherence stays close to the maximal value of $-1/2$
throughout the interaction. The maintainance of coherence is the key feature
of this level scheme, allowing for much better spin squeezing than in other
projection schemes. A similar idea was used to improve spin squeezing in
cavity-field interactions \cite{bermandantan} .

\section{Emitted field}

A wave equation for the signal field can be written where the polarization of
the medium acts as a source. Defining the positive frequency part of the
polarization resulting from a single atom (denoted by $\alpha$) located at
position $\mathbf{r}$ as $\widehat{P}_{\alpha}^{(+)}(t)=\widehat{P}_{\alpha
}(t)e^{i(k_{s}z-\Omega_{s}t)}$, its envelope is given by%
\begin{equation}
\widehat{P}_{\alpha}(t)=\hbar\left[  d_{23}\sigma_{23}(t)+h.c.\right]
\label{oneatpol}%
\end{equation}
The assumption of Fresnel number close to unity for the pencil-shaped medium
can be invoked now; this leads to a one-dimensional behavior of the
propagation of the signal field. In consequence, as in Refs. \cite{raymer,
dantan}, $z$ dependent continuous operators can be defined by performing an
average over infinitesimal slices in the transverse direction of the medium.
The continuous polarization operator is thus defined as%
\begin{equation}
\widehat{P}(z,t)=\underset{\Delta z\rightarrow0}{\lim}\frac{1}{\Delta V_{z}%
}\underset{\alpha\in\Delta V_{z}}{%
{\textstyle\sum}
}\widehat{P}_{\alpha}(t), \label{contpol}%
\end{equation}
where $\Delta V_{z}=A\Delta z$ is the volume of a slice and the sum is
performed over all atoms in the slice (number of atoms in a slice $N_{z}%
=n_{a}A\Delta z$). Continuous atomic operators can also be defined as%
\begin{equation}
\widehat{O}(z,t)=\underset{\Delta z\rightarrow0}{\lim}\frac{1}{N_{z}}%
\underset{\alpha\in\Delta V_{z}}{%
{\textstyle\sum}
}\widehat{O}_{\alpha}(t). \label{contatop}%
\end{equation}
Replacing the expression for $\mathbf{\sigma}_{23}$ previously derived in Eq.
(\ref{oneatpol}) and making use of continuous atomic operators, Eq.
(\ref{contpol}) becomes:%
\begin{equation}
\widehat{P}(z,t)=\hbar n_{a}d_{23}f(t)\left[  \sigma_{11}(z)-\sigma
_{22}(z)\right]  .
\end{equation}
where the notation $f(t)=\chi_{1}^{\ast}(t)\chi_{2}(t)\chi_{p}(t)/\Delta
\Delta_{p}\delta$ has been made$.$As in Ref. \cite{raymer, dantan}, the
quantized signal field amplitude (the positive frequency part) can be written
as $\widehat{E}^{(+)}(z,t)=\mathcal{E}_{s}\widehat{E}_{s}(z,t)e^{i(k_{s}%
z-\Omega_{s}t)},$ where $\mathcal{E}_{s}=\sqrt{\hbar\Omega_{s}/2\epsilon
_{0}AL}$ and $\widehat{E}_{s}(z,t)$ is a slowly varying envelope operator. The
wave equation in terms of slowly varying field and polarization envelope
operators can be written as%
\begin{equation}
\left[  \partial_{z}+\frac{1}{c}\partial_{t}\right]  \widehat{E}%
_{s}(z,t)=-i\left[  \frac{k_{s}}{2\epsilon_{0}}\right]  \widehat{P}(z,t).
\end{equation}
Setting $K(t)=\hbar k_{s}d_{23}f(t)/2\epsilon_{0}$, we find that population
fluctuations at each point in the medium are connected to the signal field by
\begin{equation}
\left[  \partial_{z}+\frac{1}{c}\partial_{t}\right]  \widehat{E}%
_{s}(z,t)=-in_{a}K(t)\left[  \sigma_{11}(z)-\sigma_{22}(z)\right]  .
\label{fieldeq}%
\end{equation}
If collective atomic operators are defined%
\begin{equation}
S_{z}=\frac{N_{a}}{2L}\underset{0}{\overset{L}{\int}}dz\left[  \sigma
_{11}(z)-\sigma_{22}(z)\right]  , \label{collective}%
\end{equation}
the solution for Eq. (\ref{fieldeq}) is found [for derivation see Appendix A]
to be%
\begin{equation}
\widehat{E}_{s}(L,t)=\widehat{E}_{s}(0,t)-i\frac{2K(t)}{A}S_{z}.
\end{equation}

The result states that the signal field amplitude exiting the sample is
amplified by a quantity proportional to the collective population operator.
The factor $2[K(t)/A]S_{z}$ can also be reexpressed as $n_{a}[2S_{z}/N_{a}]L$,
which shows a linear increase with the length and atomic number density of the
sample. The intensity of the emitted field can be calculated as an expectation value%

\begin{equation}
I_{s}\sim\left\langle \widehat{E}_{s}(L,t)\widehat{E}_{s}^{\dagger
}(L,t)\right\rangle =\frac{4\left\vert K(t)\right\vert ^{2}}{A^{2}%
}\left\langle S_{z}^{2}\right\rangle .
\end{equation}

An initial population imbalance in the ground substates gives rise to a signal
quadratic in the number of atoms in the sample. However, when the collective
state of the system is a coherent one, polarized along the $x$ direction, for
example, population fluctuations only are reflected in the emitted field. The
variance of $S_{z}$ is in this case $N_{a}^{2}/4,$ which leads to a gain in
the field intensity, linear in number of atoms.

\section{Effective Hamiltonian and entanglement generation}

The conditional atomic generation process is similar to an EPR-type
experiment, where entanglement is created between two subsystems (medium and
signal field, in our case) by means of an interaction that lasts a finite time
(duration of pulses); a measurement (detection of signal photon number) is
performed on one of the subsystems (field) long after the interaction has
ceased. Consequently, the other subsystem (atoms) is projected onto the state
entangled with the state indicated by the detection outcome. A wave function
approach (or density matrix, when imperfect detection is accounted for) can be
taken to describe the coupled evolution of the two subsystems, while a
continuous measurement theory is particularized to this case to describe the
nondeterministic evolution of the system during the detection process.

The results of Ref. \cite{berman-genes} are used in what follows. The complete
details of the derivation of an effective Hamiltonian are found in Appendix C
of Ref. \cite{berman-genes}, where a similar calculation is described. The
measurement process, both under the assumption of perfect detection and
including imperfect detectors, is also presented in details in Sec. V. of Ref.
\cite{berman-genes}. We are concerned here, rather with the main differences
between our proposed scheme and the ones proposed elsewhere.

The interaction between the signal field amplitude operator and the atoms in
the sample is written in the Heisenberg picture. An integration over the
transverse wave vector components of the generated field (allowed by the
assumption of Fresnel number close to unity) followed by one over $x$ and $y$
leads to a one-dimensional formulation of the problem, where the continuous
atomic operators are specified only by their $z$ spatial location, while the
field has a transverse spatial extent $A$ (matching the cross-sectional area
of the medium) [see also \cite{dantan}]. The coherence between levels $2$ and
$3$ is thereafter replaced, using Eq. (\ref{sigma23final}), to lead to an
effective Hamiltonian%
\begin{equation}
H_{eff}(t)=\hbar b(t)\left(
{\textstyle\int}
dk_{z}d_{y}^{\dag}(k_{z})e^{i(\omega_{k}-\Omega)t}\right)  S_{z}+h.c,
\label{effHam}%
\end{equation}
where $b(t)=\left[  4\pi d_{23}\mathcal{E}_{s}/\hbar\sqrt{A}\right]  f(t)$,
while $d_{y}^{\dag}(k_{z})$ is a one-dimensional field operator defined as%
\begin{equation}
d_{\lambda}(k_{z})=\frac{1}{2\pi\sqrt{A}}\underset{A}{%
{\textstyle\int}
}dxdy%
{\textstyle\int}
dk_{x}dk_{y}a_{\lambda}(\mathbf{k})e^{ik_{x}x}e^{ik_{y}y}%
\end{equation}
and satisfying the following commutation relations $\left[  d_{\lambda}%
(k_{z}),d_{\lambda^{\prime}}^{\dag}(k_{z}^{\prime})\right]  =\delta
(k_{z}-k_{z}^{\prime})\delta_{\lambda\lambda^{\prime}}$.

With the observation that the Hamiltonian commutes with itself at different
times $\left[  H_{eff}(t),H_{eff}(t^{\prime})\right]  =0$, the evolution
operator over the duration of the interaction can be expressed in a simple
form:%
\begin{equation}
U(T)=\exp\left[  -\frac{i}{\hbar}\overset{T}{\underset{0}{%
{\textstyle\int}
}}dtH_{eff}(t)\right]  .
\end{equation}
The time integral brings the Fourier components of $f(t)$ the incident pulse
field envelope $\underset{0}{\overset{T}{%
{\textstyle\int}
}}dte^{i(\omega_{k}-\Omega)t}f(0,t)\simeq F(\omega_{k}-\Omega)$. The integral
over $k_{z}$ in Eq. (\ref{effHam}) can be now represented by an effective
one-photon creation operator with carrier frequency $\Omega_{s}$ and duration
$\left(  c\Delta k\right)  ^{-1}\simeq T$ defined as
\begin{equation}
c_{y}^{\dag}=\frac{c^{1/2}}{\sqrt{\overset{T}{\underset{0}{%
{\textstyle\int}
}}dt\left\vert f(t)\right\vert ^{2}}}%
{\textstyle\int}
dk_{z}F(\omega_{k}-\Omega)d_{y}^{\dag}(k_{z}),
\end{equation}
and obeying the usual commutation relation $[c_{y},c_{y}^{\dag}]=1$. This
leads to a simple form for the evolution operator%
\begin{equation}
U(T)=\exp[-iC(c_{y}^{\dag}-c_{y})S_{z}]. \label{evOp}%
\end{equation}
with%
\begin{equation}
C=\left[  4\pi d_{23}\mathcal{E}_{s}/\hbar\sqrt{A}c^{1/2}\right]  \left(
\overset{T}{\underset{0}{%
{\textstyle\int}
}}dt\left\vert f(t)\right\vert ^{2}\right)  ^{1/2}.
\end{equation}

The atoms-signal field system starts in an initial state with state vector%
\[
\left\vert \psi(0)\right\rangle =\left\vert S_{x}=S\right\rangle _{a}%
\otimes\left\vert 0\right\rangle _{f}=\overset{S}{\underset{M=-S}{\sum}%
}A(S,M)\left\vert S,M\right\rangle _{a}\otimes\left\vert 0\right\rangle
_{f}\text{,}%
\]
where the index $a$ denotes states of the atoms, while the index $f$ denotes
states of the field. The initial state of the atoms is an eigenstate of
$S_{x}$ [operator which is defined similarly to Eq. (\ref{evOp})] with
binomial coefficients $A(S,M)=\frac{1}{2^{S}}\sqrt{(2S)!/(S+M)!(S-M)!}$ (where
$S=N_{a}/2$). After a period of coherent evolution governed by the evolution
operator $U(T)$ [Eq. (\ref{evOp})], a collapse induced by a measurement with
an outcome of $n_{m}$ photons leads to the following state vector for the
atoms (assuming 100\% detection efficiency):%
\begin{equation}
\left\vert \psi_{n_{m}}\right\rangle _{a}=\overset{S}{\underset{M=-S}{%
{\textstyle\sum}
}}\frac{A(S,M)(iCM)^{n_{m}}e^{-(CM)^{2}/2}}{\sqrt{\overset{S}{\underset
{X=-S}{\sum}}\left\vert A(S,M)\right\vert ^{2}(CM)^{2n_{m}}e^{-(CM)^{2}}}%
}]\left\vert S,M\right\rangle _{a} \label{statevectorafterdetection}%
\end{equation}

This state vector describes spin squeezed states for $n_{m}=0$ and Schrodinger
cat states for $n_{m}>0$. The results are similar to the ones presented in
\cite{berman-genes} (the reader is refered to that publication for relevant
discussions and graphs), with one important exception. In \cite{berman-genes},
the value of $C$ is limited to small values ($C\leq\sqrt{(n_{a}\lambda
^{2}L)/2N_{a}}$) to insure that spontaneous emission does not lead to mean
spin depolarization. Here, that restriction does not apply since the total
coherence is not seriously affected by spontaneous emission. In consequence,
the results obtained here overlap with the ideal case presented in
\cite{berman-genes}, where spin squeezing close to the Heisenberg limit and
well-resolved Schrodinger cat states can be obtained.

\section{Conclusions}

We presented a probabilistic scheme in which the detection of photon number
induces the collapse of the quantum state of a collection of atoms onto either
a spin squeezed or a Schrodinger cat state. The main result of the paper is
that spontaneous decay does not play a critical role in the decoherence of the
system, therefore allowing one to obtain a squeezing parameter close to the
Heisenberg limit (assuming perfect detection). This has been done by
maintaining the system in a dark state by choosing the appropriate
configuration of driving pulses.

\section{Acknowledgements}

This work is supported by the National Science Foundation under Grant No.
PHY-0244841 and the FOCUS Center grant.

\section{Appendix : Solution for the field equation}

Using the Laplace transform with respect with $z~$(defined as
$o(s)=\mathcal{L}\left\{  O(z)\right\}  =\underset{0}{\overset{\infty}{\int}%
}dzO(z)e^{-sz}$) and defining $e(s,t)=\mathcal{L}\left\{  \widehat{E}%
_{s}(z,t)\right\}  $ and $\sigma_{z}(s)=\mathcal{L}\left\{  \sigma
_{z}(z)\right\}  $ Eq. (\ref{fieldeq}) becomes%
\begin{equation}
\partial_{t}\widehat{e}_{s}(s,t)+cs\widehat{e}_{s}(s,t)=c\widehat{E}%
_{s}(0,t)-icK(t)\sigma_{z}(s). \tag{A1}\label{main equation Laplace space}%
\end{equation}
A formal integration results in%
\begin{equation}
\widehat{e}_{s}(s,t)=\widehat{e}_{s}(s,0)e^{-cst}+\underset{0}{\overset
{t}{\int}}dt^{\prime}e^{-cs(t-t^{\prime})}\left[  c\widehat{E}_{s}%
(0,t^{\prime})-icK(t^{\prime})\sigma_{z}(s)\right]  . \tag{A2}%
\end{equation}
The time dependent term $K(t^{\prime})$ contains the slow varying field
envelope, which is evaluated at time $t$ leading to%
\begin{equation}
\widehat{e}_{s}(s,t)=\widehat{e}_{s}(s,0)e^{-cst}+\frac{1}{s}\left[
1-e^{-sct}\right]  \left[  \widehat{E}_{s}(0,t)-iK(t)\sigma_{z}(s)\right]  .
\tag{A3}%
\end{equation}
Applying the inverse Laplace transform $\mathcal{L}^{-1}$, a general solution
of Eq. (\ref{fieldeq}), for arbitrary $z$ and $t$, is found%
\begin{align}
\widehat{E}_{s}^{(+)}(z,t)  &  =\widehat{E}_{s}^{(+)}(z-ct,t)h(z-ct)+\widehat
{E}_{s}^{(+)}(0,t)\left[  h(z)-h(z-ct)\right]  -\tag{A4}\\
&  -iK(t)\left[  \underset{0}{\overset{z}{\int}}dz^{\prime}\widehat{\sigma
}_{z}(z^{\prime})-\left\{  \underset{0}{\overset{z-ct}{\int}}dz^{\prime
}\widehat{\sigma}_{z}(z^{\prime})\right\}  h(z-ct)\right]  .\nonumber
\end{align}
We evaluate now the field at the sample exit ($z=L$). In the limit of long
pulses ($T\gg L/c$), the Heaviside function $h(L-ct)$ is zero for most of the
time and will therefore be ignored. In terms of the collective operator
defined in Eq. (\ref{collective}), the field becomes:%
\begin{equation}
\widehat{E}_{s}^{(+)}(L,t)=\widehat{E}_{s}^{(+)}(0,t)-i\frac{2K(t)}{n_{a}%
A}S_{z}. \tag{A5}%
\end{equation}

\end{document}